\documentclass[aps,pre,reprint,twocolumn,prx,superscriptaddress,showpacs]{revtex4}
\usepackage[utf8]{inputenc}
\usepackage[english]{babel}
\usepackage[T1]{fontenc}
\usepackage{hyperref}
\usepackage{graphicx}
\usepackage{subfig}
\usepackage{subfloat}

\usepackage{amsmath}
\usepackage{amsfonts}
\usepackage{amssymb}
\usepackage[bitstream-charter]{mathdesign}

\begin{document}
 
\title{Short relaxation times but long transient times in 
both simple and complex reaction networks}

\author{Adrien Henry}
\affiliation{INRA,  Univ Paris-Sud, CNRS, APT, UMR 0320 / UMR 8120 Génétique Quantitative et Évolution, F-91190 Gif-sur-Yvette, France}

\author{Olivier C. Martin}
\affiliation{INRA, Univ Paris-Sud, CNRS, APT, UMR 0320 / UMR 8120 Génétique Quantitative et Évolution, F-91190 Gif-sur-Yvette, France}

\begin{abstract}
When relaxation towards an equilibrium or steady state is exponential 
at large times, one usually considers that the associated relaxation time $\tau$,
\emph{i.e.}, the inverse of that decay rate, is the longest
characteristic time in the system.
However that need not be true, and in particular other times
such as the \emph{lifetime} of an infinitesimal perturbation can be much longer. 
In the present work we demonstrate that this paradoxical property can arise even 
in quite simple systems such as a chain of reactions obeying mass action kinetics. 
By mathematical analysis of simple reaction networks, 
we pin-point the reason why the standard relaxation time does not provide 
relevant information on the potentially long transient
times of typical infinitesimal perturbations.
Overall, we consider four characteristic 
times and study their behavior in both simple chains
and in more complex reaction networks taken from 
the publicly available database ``Biomodels''. In all these systems
involving mass action rates, Michaelis-Menten reversible kinetics, or 
phenomenological laws for reaction rates, we find that the characteristic
times corresponding to lifetimes of tracers and of concentration perturbations
can be much longer than $\tau$.
\end{abstract}
\maketitle

\section{Introduction}

Networks have been used to model systems involving large numbers of
components, agents, or species~\cite{Albert_Barabasi_2002}. Of 
particular interest are the effects arising in such systems either because of
out-of-equilibrium dynamics or through equilibrium 
phase transitions. Collective effects are generally associated with slow dynamics, 
\emph{i.e.}, characteristic times that are much larger than the microscopic
times associated with elementary processes. In the 
present work our focus is on the emergence of large characteristic times 
in \emph{reaction} networks close to their steady state. There are many ways to define a characteristic time in a dynamical system. The 
simplest is via the asymptotic relaxation towards the steady state~\cite{Awazu2009,Jamshidi2008}, relaxation which 
often will be exponential. If so, the amplitude of the perturbation 
or ``distance'' to the steady state will decay as $\exp( - t / \tau)$ at 
very long times, from which one then defines $\tau$ to be
the \emph{relaxation time}. Although in familiar situations $\tau$ is the longest 
characteristic time, our goal here is to investigate cases where much larger times can arise. Our study focuses on reaction networks for specificity,
but our framework is generally applicable to any system. 

Reaction networks involve species that can transform
one into another. If the species are molecular, one can get insights into 
the dynamics of the system by introducing an isotopic 
\emph{tracer} and by following in time its incorporation into the
different molecular species~\cite{Meier2011}. Assume that the reaction network 
is in contact with outside reservoirs, and let 
$t_t$ be the time the tracer takes to exit the system. Surprisingly, the mean of
$t_t$, corresponding to the lifetime~\cite{Hardt1981,Waniewski2007} of the tracer
(and sometimes called the mean residence time of the tracer),
can be much \emph{greater} than $\tau$. The object of our work is to
understand such a possibility, pointing in particular to the danger of
assuming that $\tau$ is the main and longest characteristic time in
these systems. For pedagogical reasons,
we will begin by treating one-dimensional networks because
an in-depth analytical treatment is feasible there, from which one can easily
understand the influence of network size. We will then 
study more general systems using reaction networks published
by other authors. In all cases, we compare the
behaviors of \emph{four} characteristic times in these systems, investigating
the causes that can render them non informative
or make their ratios diverge.

\section{Models and Methods}
\subsection{Networks, molecular species and associated reactions}

A metabolic network consists of a set of reactions and associated metabolites. It is convenient to represent such a network as a graph where the nodes are associated with metabolites; these are linked together by edges when there is a reaction that includes them as substrate and product. Such edges may be uni or bi-directional, accounting for the reversibility of the associated reaction. 
Let there be $N$ metabolites $M_i$ ($i=1,...,N)$ and define $C_i$ 
as the concentration of $M_i$. We are interested in the dynamics of 
the $C_i$, \emph{i.e.}, how these quantities change with time and in the 
corresponding fluxes through the different reactions. Specifically, we 
shall study the dynamics close to the system's steady state and we shall probe the
associated characteristic times. To facilitate the mathematical understanding of these times, we shall first focus on a particular kind of network consisting of a linear chain of reactions. In that situation, we order the  metabolites from $0$ to $N+1$ where the metabolite $M_i$ is the product of reaction $R_i$ whose substrate is metabolite $M_{i - 1}$:
  \begin{equation}
    \label{eq:linear_chain}
    {M_0}\overset{v_1}{\leftrightarrow} M_1 \overset{v_2}{\leftrightarrow} ... \overset{v_N}{\leftrightarrow} M_{N}\overset{v_{N+1}}{\leftrightarrow} {M_{N+1}}
  \end{equation}

The metabolites ${M_0}$ and ${M_{N+1}}$ reside in infinite reservoirs at the two extremities of the chain so their concentrations are constant. By convention, the forward direction in such a chain goes from $M_0$ to $M_{N+1}$. Once understood the characteristic times in this system, we shall use the insight thereby gained to probe the situation in more realistic metabolic networks having branches and loops.

Reactions transform metabolites into other metabolites but it is necessary still to specify the actual kinetics. When a reaction happens spontaneously, without the need for a catalyst, it can be modeled by a mass action rate law (MA) where the flux is given 
by
  \begin{equation}
    \label{eq:MA_rate_law}
    v^{MA}_i = a_iC_{i - 1}  - b_iC_{i} \, .
  \end{equation}
To be specific, one can consider using the usual convention whereby concentrations 
are measured in Moles per liter and fluxes in Moles per liter per second. The parameter $a_i$ (resp. $b_i$) is then the probability per second that a molecule of metabolite $M_{i - 1}$ (resp. $M_{i}$) spontaneously transforms into a molecule of metabolite $M_i$ (resp. $M_{i - 1}$). Note that Eq.~\ref{eq:MA_rate_law} gives the total
flux which is the forward flux minus the backward flux. 

In practice, one is often interested in catalyzed reactions where the spontaneous rates are terribly low. For instance, in biochemistry, most reactions are catalyzed by enzymes; the catalysis allows for rates that can be enhanced by a factor of $10^{10}$ or more. For any such enzymatic reaction, the rate may be limited by the amount of enzyme and is no longer directly proportional to metabolite concentration. Generally, the relation between substrate concentration and reaction rate grows linearly at low concentrations  and then saturates at high concentrations of substrate. The reaction kinetics in this situation are typically modeled by the so called reversible Michaelis-Menten-Henri ($MMH$) law~\cite{haldanes}. In the case of a reaction involving one substrate and one product, the flux is given by
  \begin{equation}
    \label{eq:MMH_rate_law}
    v^{MMH}_i = \frac{  \alpha_i \frac{C_{i - 1}}{K_S^i}  - \beta_i \frac{C_i}{K_P^i}}{1 + \frac{C_{i - 1}}{K_S^i} + \frac{C_i}{K_P^i}}  \, .
  \end{equation}
Here, $ \alpha_i$ is the maximum rate in the forward direction, reached
when the substrate is in large excess and the product is absent. Similarly,  
$\beta_i$ is the maximum rate in the backward direction. The maximum forward rate is proportional to the enzyme concentration and is often decomposed as $ \alpha = k_{cat}E$ with $E$ being the enzyme concentration and $k_{cat}$ the maximum number of reactions catalyzed by one molecule of enzyme per unit of time. $K_S^i$ and $K_P^i$, called the Michaelis constants respectively for substrate and product, are characteristic concentrations which set the scale for when the reaction becomes saturated in substrate or in product. For a $MMH$ reaction in the absence of the product, $K_S$ is the concentration for which the rate is at half of its maximum value.

\subsection{Determining steady states}
When a physical system is not driven by outside forces, it goes to its equilibrium state where all net reaction fluxes are 0. In the context of our one dimensional model, that can only arise if the free energies of the two reservoirs are equal, corresponding to tuning the concentrations so that their ratio is the equilibrium one. Outside of that special case, the system will be out of equilibrium and concentrations will change in time until a steady state is reached which necessarily will have non zero fluxes. This steady state is generally unique if there are no regulatory processes but for our study to be completely general, we will not assume uniqueness of the steady state, we shall simply consider a stable steady state and investigate its characteristic times.

We have followed two approaches for 
determining steady states (leading to identical results):
\begin{enumerate}
\item solve the steady state equations $dC_i /dt = 0$ which
we do numerically using root finding (routine ``find-root'' 
in Python). For any given boundary conditions,
\emph{i.e.}, concentrations $C_0$ and $C_{N+1}$, this leads to 
a list of steady-state concentrations $\overrightarrow{C^{ss}}$. It is
necessary to check that the resulting steady state is linearly stable. This 
check can be performed using the linearized equations about the steady state. If
$\overrightarrow{\delta C}$ is the (infinitesimal) difference
between the actual concentrations and those in the
steady state, one has
\begin{equation}
\frac{d \overrightarrow{\delta C}}{dt} = \bold{J^{(c)}} \overrightarrow{\delta C}
\end{equation}
\begin{equation}
\bold{J^{(c)}}_{ij}=\begin{cases}
A_i \quad\mbox{if}\quad j=i - 1\\
- (A_{i+1}+B_i) \quad\mbox{if}\quad j=i\\
B_{i+1} \quad\mbox{if}\quad j=i+1\\
0\quad\mbox{otherwise} \end{cases} \, 
\label{eq:linearized_dynamics_concentration}
\end{equation}
where the $A_i$ and $B_i$ are related to the terms entering
Eq.~\ref{eq:MA_rate_law} for mass action and Eq.~\ref{eq:MMH_rate_law} for Michaelis Menten Henri
as specified in Table~\ref{tab:AB_parameters}. 
$\bold{J^{(c)}}$ is the $N \times N$ Jacobian matrix with indices 
$i$ and $j$ going from 1 to $N$; the superscript $c$ 
refers to the fact that it describes the (linearized) dynamics
of (perturbed) \emph{concentrations}. The steady state is stable if all
the eigenvalues of the Jacobian have negative real part.
\item follow the concentrations using the dynamical equations
(the system of ordinary differential equations specified by 
the kinetic laws) and extract
the long time limit of the concentrations. This requires
extrapolation, but generically takes one to a stable steady
state. 
\end{enumerate}

\subsection{Defining four characteristic times}
\begin{itemize}
\item The first characteristic time is the 
\emph{relaxation} time 
defined as $ - 1/\lambda_1^{(c)}$ where $\lambda_1^{(c)}$ is the real part of the
leading eigenvalue of $\bold{J^{(c)}}$ having the largest real part. 
Because this time is defined via the linearized dynamics for the 
\emph{concentrations} about the steady state, we shall refer to it as $\tau_c$.

\item The second characteristic time is
the previously mentioned tracer \emph{lifetime} (or mean residence time), 
which we denote by $T_t$. The motivation for introducing this quantity comes
from tracer experiments in chemical networks where isotopic labels are used
to follow atoms as reactions progress. Instead of introducing a perturbation
to concentrations, this approach labels (e.g. via an NMR pulse) atoms of one 
metabolite $M_k$ at $t=0$ without changing any
concentrations. In practice this labeling affects only a fraction
of the molecules. The effect of this labeling is
to leave the fluxes unperturbed as well. The
system stays in its steady state, it is just that some of these concentrations become labeled. Note that when
one labeled metabolite is tranformed into another, the labeling follows
because the labeled atoms.

Let us study the time evolution of the 
concentrations of these tracers 
$\overrightarrow{ C_t}=\{  C_{t,1},
  C_{t,2}, \ldots,  C_{t,N}\}$ (the
subscript $t$ is for \emph{tracer}).
As previously introduced, let 
$\overrightarrow{C^{ss}}=\{C^{ss}_{1}, C^{ss}_{2}, \ldots, C^{ss}_{N}\}$
be the steady state concentrations. Consider the
reaction $R_i$ and let ${\phi^f}_i$ be its forward flux and 
${\phi^b}_i$ its backward flux in the steady state. Then the labeled concentration
$ C_{t,i}$ will include an incoming term given by the rescaled forward flux 
${\phi^f}_i\, C_{t,i - 1} / C^{ss}_{i - 1}$ because all metabolite
molecules (labeled or not) have an equal probability of 
participating in the reaction $R_i$. As a result, the dynamics of the
tracer concentrations is 
\begin{equation}
\frac{d \overrightarrow{C_t}}{dt} = \bold{J^{(t)}} \overrightarrow{ C_t}
\end{equation}
\begin{equation}
\bold{J^{(t)}}_{ij}=\begin{cases}
\phi^f_{i} / C^{ss}_{j} \quad\mbox{if}\quad j=i - 1\\
-(\phi^f_{i} / C^{ss}_{i - 1} + \phi^b_{i - 1} / C^{ss}_{i - 1} ) \quad\mbox{if}\quad j=i\\
\phi^b_{i} / C^{ss}_{i}  \quad\mbox{if}\quad j=i+1\\
0\quad\mbox{otherwise} \end{cases} 
\end{equation}
Note that these linear dynamics are exact even if $C_{t,i}$ is not infinitesimal.
In general, the matrix $\bold{J}^{(t)}$ has no reason to be identical 
to $\bold{J^{(c)}}$. By
exponentiating, one has the expression for the labeled concentrations
at all times: 
$\overrightarrow{ C_t}(t) = \exp( t \bold{J^{(t)}}) 
\overrightarrow{ C_t}(0)$. The lifetime $T_t$ of the tracer is 
then obtained as the
average over time of the survival probability:
\begin{equation}
T_t = \frac{\int_0^{\infty} |\overrightarrow{ C_t}(t) | dt}{|\overrightarrow{ C_t}(0) | }
\label{eq:def_Tt}
\end{equation}
In this equation, $|\overrightarrow{ C_t}(t) |$ is the norm of the
vector. For our study, we use the 
$L_1$ norm because it makes more sense for an atomic tracer
which is conserved. Note also that $T_f$ in Eq.~\ref{eq:def_Tt} is 
the direct analog of the mean lifetime of a decaying positive
\emph{scalar} quantity; the norm allows one to extend
the notion to a vector in a straightforward manner.

\item The previous definition of lifetime of a tracer can be generalized to
the lifetime of any quantity and in particular to
a perturbation to steady-state concentrations. Suppose one 
introduces at $t=0$ an infinitesimal perturbation in the
concentrations, $\overrightarrow{\delta C}(0)$. Then according
to Eq.~\ref{eq:linearized_dynamics_concentration}, 
$\overrightarrow{\delta C}(t) = \exp( t \bold{J^{(c)}}) \overrightarrow{\delta C}(0)$. In direct analogy
with Eq.~\ref{eq:def_Tt}, the lifetime of that perturbation is
\begin{equation}
\label{eq:def_Tc}
T_c = \frac{\int_0^{\infty} |\overrightarrow{\delta C}(t) | dt}
{|\overrightarrow{\delta C}(0) |}
\end{equation}
providing a third characteristic time of our system, referred to as
the lifetime of a concentration perturbation. 
To be completely general, 
both here and for the tracer lifetimes, the vectors of
concentrations should be actually the deviations of their values from
their long time limit. Indeed, if there were no reservoir and thus no
exit possible of the atoms, the long time limit of the perturbation or tracer
concentration would not be 0.

\item Our fourth and last characteristic time is $\tau_t$, defined as
$-1/\lambda_{1}^{(t)}$ where $\lambda_{1}^{(t)}$ is here the real part of the
leading eigenvalue of $\bold{J^{(t)}}$. It corresponds thus to the usual
relaxation time but for the tracer molecules rather than for the  
metabolite concentrations, thus the subscript $t$.

\end{itemize}

\section{Behavior of characteristic times in the one-dimensional network}

As can be seen from the four characteristic times defined
in the previous section, we distinguish two 
properties of a metabolic system: (i) the dynamics of an infinitesimal perturbation
in the concentration of metabolites and (ii) the spreading and drift of tracers. 
Each of these properties can be considered when reaction kinetics
are given by $MA$ or $MMH$ rate laws. In each case one can define
both the standard relaxation time based on the asymptotic
decay rate and a lifetime which measures the characteristic
time needed for the system to return close to its steady state.
In the case of a chain of reactions with the same kinetic parameters,
the homogeneity allows us to obtain results analytically. 
For instance in the case of $MA$, the linearized dynamics
($\bold{J^{(c)}}$ and $\bold{J^{(t)}}$) are independent of the steady state
chosen (that is the concentrations of $M_0$ and 
$M_{N+1}$ do not enter) and the matrices are sufficiently simple
for one to obtain in closed form the eigenvectors and
eigenvalues. In the case of a $MMH$ 
framework, when one performs the linearization about the steady state, 
the resulting system is homogeneous only if the steady state itself
is homogeneous, which requires that all the metabolites have the same concentrations. 
When this is 
the case, the steady state is again obtained in closed form. Furthermore, 
the eigenvectors and eigenvalues can be derived analytically, which gives
us then the formulas for $\tau_c$ and $\tau_t$. Unfortunately the
study of the lifetimes $T_c$ and $T_t$ requires resorting to numerical
methods, but these are relatively straightforward as they reduce
to calculating exponentials of the matrices $\bold{J^{(c)}}$ and $\bold{J^{(t)}}$
and performing the integrations in Eq.~\ref{eq:def_Tc} and~\ref{eq:def_Tt}.  For the initial perturbation, for simplicity
we take $\overrightarrow{\delta C}(0)$ and 
$\overrightarrow{C_t}(0)$ to vanish everywhere 
except on the site at the center of the chain where the value is set to 1. 
For an even number of sites, there is no such center so we average 
over the two most central sites.

\subsection{Long transient times drive the gap between lifetimes 
and relaxation times}
\begin{figure}
  \centering
\includegraphics[width=\columnwidth]{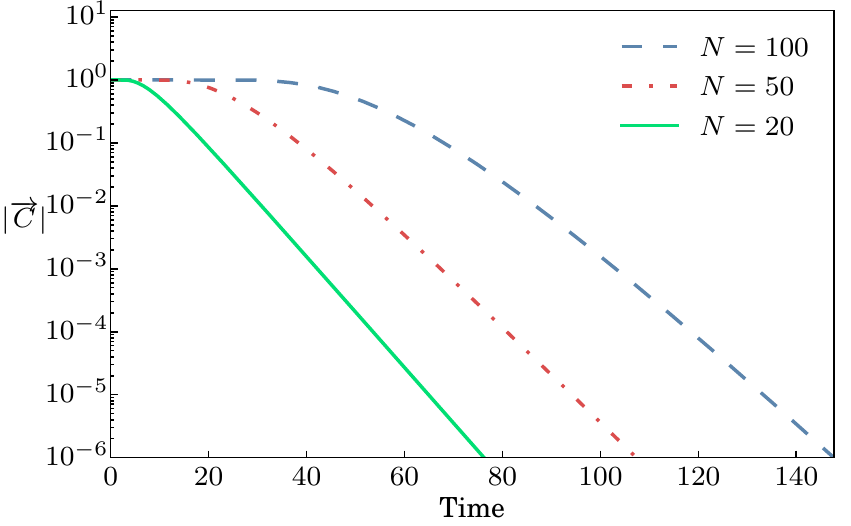}
  \caption{Decrease with time of $|\protect \overrightarrow{\delta C}|$ (or $|\protect \overrightarrow{C_t}|$) for an initial $t=0$ excess concentration  (or labeling) localized at a site in the middle of the chain of reactions. The y axis is on a log scale so that one can
see the asymptotic exponential decay as a straight line of
slope $ -1/\tau$: $\tau_{20}=4.92$, $\tau_{50}=5.65$, and $\tau_{100}=5.78$. All $N$ mass action reactions have $a=2$ and $b=1$.
Shown are cases with $N=20, 50$ and $N=100$. 
}
\label{fig:decay_tracer}
\end{figure}

The integral in Eq.~\ref{eq:def_Tt} depends on
$\overrightarrow{ C_t}(t) = \exp( t \bold{J^{(t)}}) \overrightarrow{ C_t}(0)$
which can be written using spectral decomposition
as a sum of $N$ terms, each term being associated
with one eigenmode and having the time dependence 
$\exp( t \lambda_i^{(t)})$ where $\lambda_i^{(t)}$ is the associated eigenvalue.
When $N=1$, $\overrightarrow{ C_t}(t)$ is a constant
times a single decaying exponential. Plugging into Eq.~\ref{eq:def_Tt}
then reveals that $T_t=\tau_t$. The paradox whereby
$T_t$ can be much larger than $\tau_t$ arises only when 
$N \gg 1$. It is true that each of the $N$ terms contributing
to the spectral decomposition of $\overrightarrow{ C_t}(t)$
decays in magnitude at least as fast as $\exp( - t / \tau)$ but that does \emph{not} 
mean that the sum of these terms has that behavior on time
scales comparable to $\tau$. Indeed, the terms are not all of the
same sign, and their cancellations can lead to long transients
before the asymptotic behavior (the exponential decay) 
prevails. To illustrate this,
we show in Fig.~\ref{fig:decay_tracer} the $L_1$ norm of 
$\overrightarrow{ C_t}(t)$ as a function of $t$ in our toy model consisting of 
a chain with $a$'s and $b$'s identical across $MA$ reactions. 
At large times, one sees the exponential decay (a straight line
on this semi-log plot) but this asymptotic behavior may set in at times
only much longer than $\tau$ itself. The cancellation at short times
just mentioned is particularly striking: the curve is very flat 
for a very long time before it begins to decrease. That waiting
time contributes to the large difference between $T_t$ and
$\tau_t$ and is associated with the transient time one must wait for tracer 
molecules to exit the system.
\begin{figure}
  \centering
\includegraphics[width=\columnwidth]{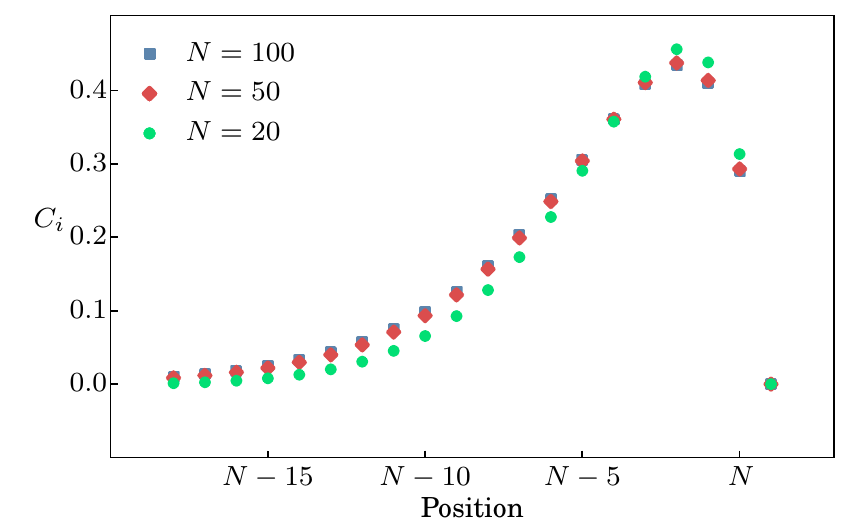}
\caption{Leading eigenmode profile for the 20 last metabolites of the chain. Parameters: $a=2$, $b=1$, and $N=20,50,100$.}
\label{fig:eigenvector}
\end{figure}

\subsection{Dependence of the characteristic times on $N$}
\begin{figure*}
  \centering
  \subfloat{\includegraphics[width=\columnwidth]{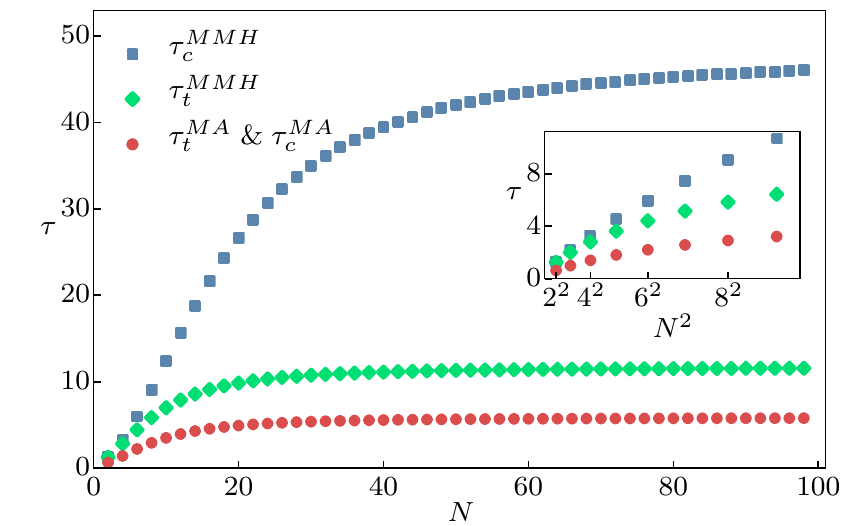}}
  \subfloat{\includegraphics[width=\columnwidth]{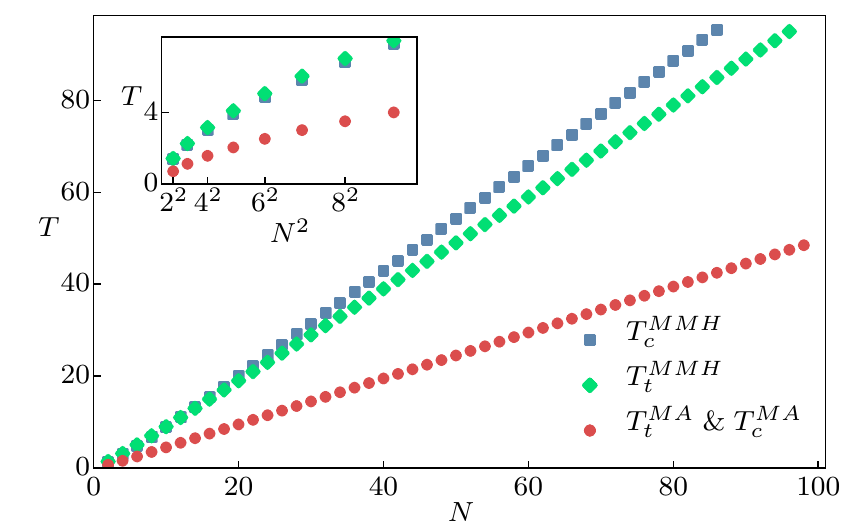}}
  \caption[width=\linewidth]{Relaxation times(\emph{left}) and lifetimes(\emph{right}) for chains between 2 and 100 metabolites long for a perturbation of concentrations and a tracer using the mass action or the Michaelis-Menten-Henri framework. Parameters: $a=2$ and $b=1$, $K_S = K_P = 2$ and $\alpha = aK_S$ and $\beta = bK_P$ so that the three conditions are comparable. The large $N$ relaxation times are respectively $\tau_{c,\,lim}^{MA} = \tau_{t,\,lim}^{MA} = 5.83$, $\tau_{t,\,lim}^{MMH}=11.66$, $\tau_{c,\,lim}^{MMH}=47.66$. The transition sizes between a quadratic and constant or linear behavior are $N_{c,\,cross}^{MA} = N_{t,\,cross}^{MA} = 7$, $N_{t,\,cross}^{MMH}=7$, $N_{c,\,cross}^{MMH}=17$.}
  \label{fig:relaxation_times}
\end{figure*}

Assuming the reactions to all have the same parameters and that 
the steady state is also homogeneous (cf. previous remarks), 
the relaxation time (be it $\tau_c$ or $\tau_t$) can be obtained by using the translation invariance  of $\bold{J^{(c)}}$ and $\bold{J^{(t)}}$. Each eigenvector is a product of a sine and an exponential. The formula for the eigenvalues leads to
\begin{equation}
  \label{eq:tau_analytical}
  \tau = \frac{1}{A+B  - 2\sqrt{A\,B}\cos \left( \frac{\pi}{N+1} \right) }
\end{equation}
where the quantities $A$ and $B$ are the forward and backward probability of transition per unit of time in the equations linearized about the steady state, entering in 
$\bold{J^{(c)}}$ for $\tau_c$ and in $\bold{J^{(t)}}$ for $\tau_t$. They 
depend on whether one considers $MA$ or $MMH$ reaction 
kinetics and whether one considers a concentration perturbation or a 
tracer, the different cases being enumerated in Table~\ref{tab:AB_parameters}.
\begin{table}[h!]\centering
  \begin{tabular}{c|cccc}
    Parameter & $MA-c$ & $MA-t$ & $MMH-c$ & $MMH-t$\\\hline
    $A$ & $a$ &  $a$ & $\frac{ \alpha -F}{K_SS}$ & $\frac{ \alpha}{K_SS}$ \\
    $B$ & $b$ & $b$ & $\frac{ \beta+F}{K_PS}$& $\frac{ \beta}{K_PS}$ \\
  \end{tabular}
\caption{Value of the $A$ and $B$ parameters for the four situations considered. $F$ and $S=(1 + c^{ss}/K_S +c^{ss}/K_P)$ are respectively the flux and the saturation factor at steady state in the network for the reactions, the 
system being by hypothesis homogeneous.
The ``c'' (respectively the ``t'') appended to $MA$ and $MMH$ denotes that it is the 
perturbed concentrations (respectively the tracer concentrations) that are concerned.}
\label{tab:AB_parameters}
\end{table}

The $\tau$s in the four cases are given by a standardized formula
(Eq.~\ref{eq:tau_analytical}), it is just that
the proper $A$ and $B$ coefficients must be used. Note that for $MA$ kinetics,
$\bold{J^{(c)}} = \bold{J^{(t)}}$ so $\tau_c = \tau_t$. Furthermore, in both $MA$ and $MMH$ frameworks, when the relative difference
between $A$ and $B$ is small, the $\tau$s exhibit two 
different regimes, one for small chains and one for long chains. For 
a short chain, $N \ll N^{cross} = \frac{2B\pi}{A - B}$, the characteristic 
times \emph{grow quadratically} with the number of metabolites in the chain, a feature 
characteristic of diffusing systems for the simple reason
that if $A=B$, the dynamics is purely diffusive. When 
$N$ is much above this crossover value, $\tau_c$ and $\tau_t$ 
become independent of the chain length as can be seen directly 
by setting to 1 the cosine in Eq.~\ref{eq:tau_analytical}. 

Note that the crossover size $N^{cross}$ diverges as the inverse of $A - B$. Furthermore, in the context of $MMH$ reaction kinetics, 
this crossover occurs for larger chain lengths when considering the 
dynamics of a concentration perturbation than when considering
tracers because the 
\emph{saturation} has the effect of reducing the difference
between $A$ and $B$. 
To illustrate these effects, we display
in Fig.~\ref{fig:relaxation_times}
the relaxation times as a function of the chain length $N$
for particular values of the kinetic parameters.
As for $MA$, $\tau_c$ and $\tau_t$ do not
increase asymptotically with $N$, the characteristic
times become  independent of the system size. 
To understand how this occurs, let us examine the leading eigenvector. 
Its entries depend exponentially on the index $i$ of the node and so its profile
is biased towards the largest indices. If the eigenvector with the
largest eigenvalue becomes dominant, the major part of the deviation 
from the steady state is located on a few 
metabolites (about $N^{cross}$) at the end of the network. 
As illustrated in Fig.~\ref{fig:eigenvector}, if one increases the number of metabolites, that eigenmode just gets shifted to stay at the same position when measured from the end
of the chain. As a consequence, increasing $N$ does not affect the corresponding
eigenvalue which determines $\tau$. Thus $\tau_c$ and $\tau_t$ 
become independent of $N$ at large $N$.

For the $T_c$ and $T_t$ lifetimes, we did not derive a closed form expression
but one can still distinguish between two regimes. If $A - B$ is small
the behavior for small $N$ is diffusion-like so $T_c$ and $T_t$
increases quadratically with $N$. In contrast, for long chains, if
$A \ne B$, one has a regime where $T_c$ and $T_t$ grow
linearly with $N$. Similar arguments as for the 
relaxation times $\tau$ can be invoked to 
explain these two regimes. In small networks, the diffusion to the two sides of the chain dominates 
over the mean drift toward one end of the chain. In large networks, 
assuming $A > B$, most of the transient time dominating $T_c$ and $T_t$ 
is dedicated to the transport 
of the molecules to the $N+1$ end, therefore 
that transient time is roughly equal to $N$ divided by the drift 
velocity (which is proportional to $(A-B)$). We illustrate
these behaviors in Fig.~\ref{fig:relaxation_times} where one sees
again that the various cases behave similarly with the network length.
(We already noted that for $MA$ kinetics,
$\bold{J^{(c)}} = \bold{J^{(t)}}$;
as a consequence one has $T_c = T_t$
there, just as one has $\tau_c = \tau_t$.)

\subsection{Effect of the saturation on the characteristic times}

The major differences between $MA$ and $MMH$ come from the effect of the saturation. In the case of the $MA$ rate laws, there is no saturation while saturation effects can be important in $MMH$ kinetics. This difference can lead to much larger characteristic time scales in $MMH$ than in $MA$ whenever the concentrations are larger than $K_S$ or $K_P$. Furthermore, for highly saturated enzymes, the characteristic times can be very different depending on whether one observes a tracer or a perturbation of concentration.
Consider a reaction that is near saturation. Introducing a perturbation in the substrate will not change much the flux of that reaction and as a result it will take a long time to dissipate the perturbation away. On the other hand a tracer is essentially unaffected by saturation effects. Indeed, 
it is not because the reaction is saturated that the tracers cannot participate in the 
reactions. In effect, the tracers freely pass reactions that are saturated. The main consequence of this phenomenon is that 
in $MMH$ $\tau_c$ can be much larger than
$\tau_t$ (and $T_c$ can be much larger than $T_t$).

To investigate quantitatively this phenomenon of particular relevance when 
interpreting kinetic properties from tracer measurements, let us increase saturation effects by reducing $K_S$. $K_P$ could also have been reduced, but when doing so, the flux in the network may reverse which unnecessarily
complicates the analysis. Using the parameters of Table~\ref{tab:AB_parameters} 
in Eq.~\ref{eq:tau_analytical} for small values of $K_S$ gives the following analytical 
values for the dominant terms of the two relaxation times associated with a tracer ($\tau_t$) and with
a concentration perturbation ($\tau_c$):
\begin{subequations}
\label{eq:smallKS_relaxation}
  \begin{eqnarray}
    \tau_t &\approx& \frac{1}{ \alpha} \label{eq:smallKS_relaxation_tauM} \\ 
    \tau_c  &\approx& \frac{ \left(  \alpha + 2\frac{ \alpha + \beta}{K_P} - 2 \sqrt{\frac{ \alpha}{c}\frac{ \alpha+\beta}{K_P}\left(1+\frac{ \alpha+\beta}{K_P}\right)}\cos 
\left( \frac{ \kappa \pi}{N+1}\right)\right)^{ - 1}}{K_S} \label{eq:smallKS_relaxation_tauP} 
  \end{eqnarray}
\end{subequations}

\begin{figure}
  \centering
\includegraphics[width=\columnwidth]{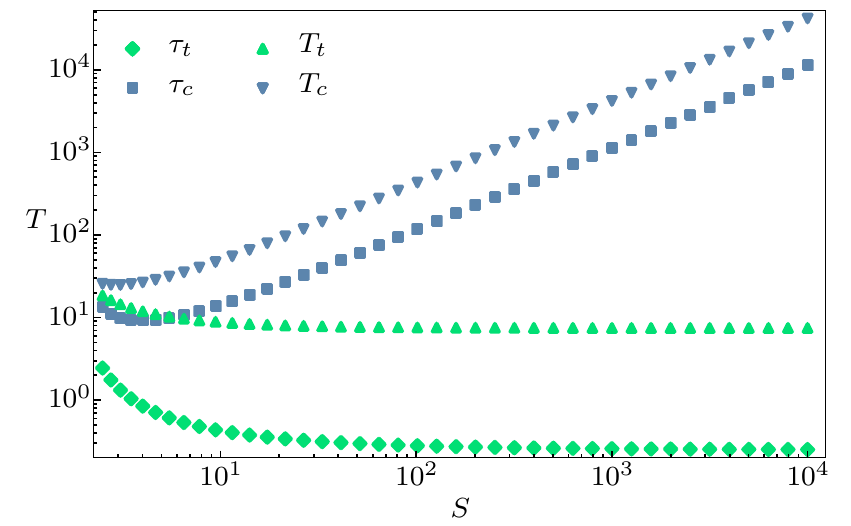}
  \caption{Relaxation times and lifetimes as a function of the saturation. Parameters: $\alpha = 4$, $\beta = 2$, $K_P = 2$, $N=30$. To vary the saturations, the parameter $K_S$ is changed over a range going from $1$ to $10^{-4}$.}
\label{fig:time_vs_sat}
\end{figure}

We see from these equations that $\tau_t$ is independent of the 
saturation while $\tau_c$ behaves linearly with $1/K_S$. Note
that the saturation $S = 1 + c^{ss}/K_S +c^{ss}/K_P$ scales
in the same way for small $K_S$. In Fig.~\ref{fig:time_vs_sat} 
we show the dependence of the $\tau$s and the
$T$s on the saturation $S$ for both a tracer and a concentration 
perturbation, assuming $MMH$ rate laws. Not surprisingly,
$T_c$ is strongly affected by $S$, just as $\tau_c$ is.

\section{Behavior of characteristic times in more general metabolic networks}

\subsection{Effects of disorder in the one dimensional chain}
In the disordered (\emph{i.e.}, heterogeneous) case we now consider,
the rates ``$a$'' and ``$b$'' for the different reactions
are taken to be independent random variables. Because every
rate is a positive variable, we draw it from a lognormal 
distribution, \emph{i.e.}, the natural logarithm of a rate $r_i$ is
distributed according to a Gaussian of mean $\mu$ and standard deviation $\sigma$. 
Consequently, the mean of $r_i$ is $\bar{\mu} = \exp(\mu + \sigma^2/2)$ and its
variance is $\bar{\sigma}^2 = ( \exp (\sigma^2) - 1) \exp(2 \mu + \sigma^2)$.
We impose $\bar{\mu}$ to be equal to the value of the rate in the homogeneous case. 
An appealing feature of that way of introducing disorder is that the 
mean drift velocity of a marked molecule in Mass Action remains unchanged, being
equal to its disorder average, $\langle a_i - b_i \rangle$. We are then left with the parameter $\bar{\sigma}$
which can go from 0 to $\infty$ and quantifies the intensity of the disorder. 
In practice, we use the same coefficient of variation for the ``on'' 
and the ``off'' reaction rates, corresponding to a single measure of
intensity of disorder: $CV = \bar{\sigma}_a/a = \bar{\sigma}_b/b$.

\begin{figure*}[ht]
  \centering
  \subfloat{\includegraphics[width=\columnwidth]{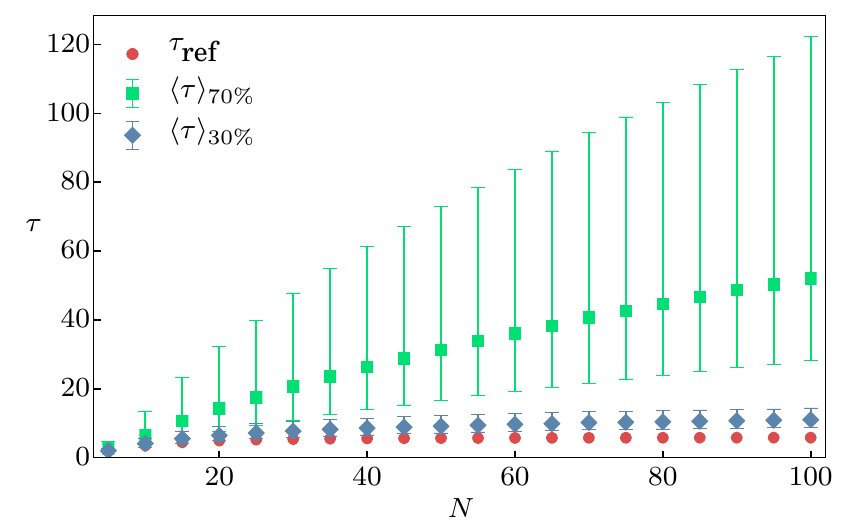}}
  \subfloat{\includegraphics[width=\columnwidth]{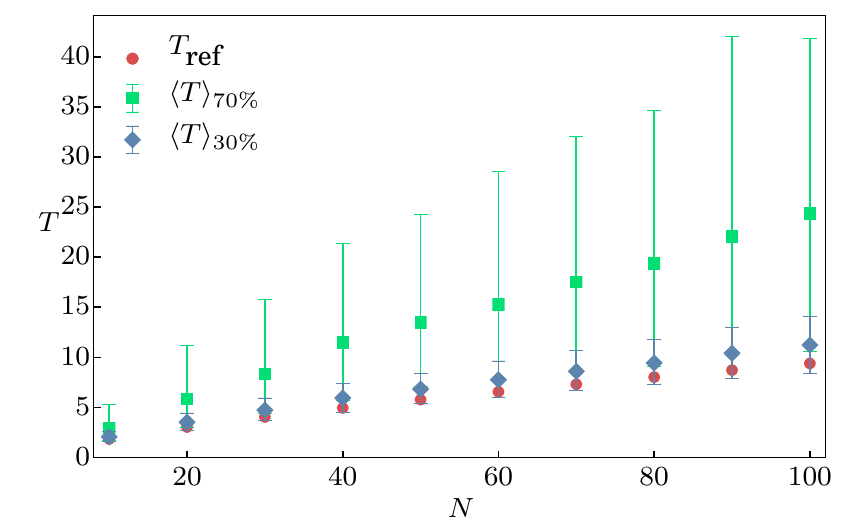}}
  \caption[width=\linewidth]{Relaxation(\emph{left}) and transit(\emph{right}) times as a function of $N$ for several intensities of disorder in
the reaction rates as measured by their coefficient of variation $CV$.
Main figure: $T=T_c=T_t$ in the case of Mass Action kinetics. The error bars show  the $68.2\%$ confidence interval, value motivated by taking one standard deviation on both sides of the mean of a Gaussian distribution. Parameters: $a=2$, $b=1$, $CV=30\%$ and $70\%$.}
  \label{fig:times_disorder}
\end{figure*}

For weak disorder, one expects little change in the values of the characteristic
times ($\tau_c, \tau_t, T_c, T_t$) compared to the homogeneous case.
However, as disorder ($CV$) increases, the 
characteristic times typically increase. To identify the typical behavior,
we have determined these characteristic times for $10,000$
realizations of the disorder and calculated the median times. We illustrate 
the associated results in 
Fig.~\ref{fig:times_disorder} for $\tau_c$ and $\tau_t$ in the case of Mass Action where those two quantities are equal. Increase is relatively mild (cf. the scales) at low $CV$ butis more marked when $CV$ is larger than $30\%$. 

Consider now the effects of disorder on the lifetimes. 
In Mass Action, $T_c = T_t$, even in the presence of disorder.
We display in Fig.~\ref{fig:times_disorder} the dependence of these quantities on $N$
for several values of $CV$ and see that disorder has little
effect as long as $CV$ is small. This can be 
justified by noticing that the drift
velocity of a molecule at site $i$ 
is $a_i - b_{i-1}$ and its ensemble average (as in an 
annealed approximation) is 
the same as without disorder, namely $a-b$. At large disorder this
argument fails because the quenched and annealed averages
are very different. An extreme case can be seen from the 
fact that a large value of ``$a$'' at
one site cannot compensate a small value at another site.
At large $CV$, one sees clear effects of disorder.
The reason should be clear: $T_c$ and $T_t$ are sensitive to 
unfavorable reactions (for instance
where $a$ is small) throughout the whole chain of reactions. 
\subsection{Networks with branches and loops}

Although quite a few biosynthetic pathways include successive
steps forming a chain of enzymatic reactions, the one dimensional systems
considered so far remain toy models because in all known organisms, large scale
biochemical metabolic networks have numerous
branches and loops. It is thus necessary to consider how characteristic time
scales might be affected by such structures. Rather than produce
artificial networks including those features, it is more relevant to
study directly the various kinetic models of metabolism
that have been proposed in the literature. The repository
``Biomodels''~\cite{BioModels2006,hucka2003systems} provides the gold standards for such models both because
the models must past tests to be deposited and because their availability
ensures that they can be compared to state of the art. Focusing further on those models
that have been manually curated, we are left with only a handful of
cases. The reason is that measuring kinetic constants of enzymes is a
very difficult task so almost always when building a kinetic model the
modeler has to use indirect methods to overcome the problem of dealing with
many unknown parameters. We studied four of these models, published
respectively in~\cite{ Teusink2000, Chassagnole2002, Mosca2012,Curto1998}.

For each of those four kinetic models, we first downloaded its SBML specification~\cite{hucka2003systems} from the
repository and exported the ordinary differential equations into Python code that can be processed. Once in our format, we determined the steady state
of the network of reactions and we then computed the matrices $\bold{J^{(c)}}$ and $\bold{J^{(t)}}$.
The associated leading eigenvectors and eigenvalues were obtained using the inverse power method,
thereby providing the values of $\tau_c$ and $\tau_t$. Furthermore numerical integration
was used to compute $T_c$ and $T_t$ according to Eqs.~\ref{eq:def_Tt}and~\ref{eq:def_Tc}.
The initial perturbation was taken to be localized at the first metabolite
produced from the compound entering the network from the outside reservoir.

In Table~\ref{tab:biomodels_results} we provide the values of the four characteristic
times for each of the Biomodels studied. The first model~\cite{Teusink2000}
contains the reactions for glycolysis in
S. cerevisiae (baker's yeast). It has 17 reactions, mostly of the reversible $MMH$ type,
and there are 14 internal metabolites. Glucose is an external metabolite
which enters the metabolism and then gets transformed. A total of
3 compounds can be excreted, all irreversibly. The characteristic
times of this model are short, from a few seconds to a few minutes. Further inspection
shows that the ordering of these four values follows the same pattern as in our
one dimensional toy model, namely
\begin{equation}
\tau_t < \tau_c < T_t < T_c
\label{eq:hierarchy}
\end{equation}
This can be justified as follows. First, $\tau_t > \tau_c$ and $T_t > T_c$
because a labeled atom is
not subject to Michelis-Menten saturation effects. The saturation of flux
in a reaction may prevent
a concentration fluctuation from being evacuated but it will not prevent labeled
atoms from going through (participating to the flux). Furthermore, in our toy model,
the $\tau$s are relatively insensitive to processes inside the network, they
depend mainly on reactions close to the excreted metabolites, while the $T$s
depend on drift throughout the whole network and thus should be larger
than the $\tau$s.

The other models follow quite closely this same pattern (cf. Table~\ref{tab:biomodels_results}). Model 2 contains the reactions for the glycolysis and the pentose phosphate pathway in \textit{E. coli}~\cite{Chassagnole2002}. It has 48 reactions and 17 internal metabolites, but we needed to remove the model's explicit time dependence to obtain a meaningful steady state. The main difference with the model 1 is the modelled organism and the glucose steady state uptake rate ($3.1~\mu mol.s^{-1}.L^{-1}$ compared to $1.5~mmol.s^{-1}.L^{-1}$) but Eq.~\ref{eq:hierarchy} is respected. 
Again model 3 contains the glycolysis and the pentose phosphate pathway, but for a human cancer cell. It has 29 reactions and 34 internal metabolites. The glucose uptake, expressed per gram of cell dry weight ($0.17mmol.s^{-1}.gcdw^{-1}$), cannot be compared to the two previous uptakes but Eq.~\ref{eq:hierarchy} is mostly verified. 

Model 4 contains the reactions for the biosynthesis of purines in \textit{E. coli}~\cite{Curto1998}. It has a total of 29 reactions and 18 internal metabolites. The main difference compared to the
other three models is that the formalism uses kinetics that are neither $MA$ nor $MMH$: the
forward and backward rates of the reactions are fractional powers of the concentrations
of the metabolites. Such fractional powers are often used phenomenologically to parametrize
allosteric or regulatory effects; they have the drawback that the flux may rise
very steeply when starting with low concentrations; although this may
be the case for some regulatory processes, it can lead to a situation where a
concentration perturbation will be evacuated more efficiently than
a labeled atom. Such a possibility seems to be realized in this model as
in Table~\ref{tab:biomodels_results} one sees that $\tau_c < \tau_t$ and $T_c < T_t$.
\begin{table}[h!]\centering
  \begin{tabular}{c|cccc}
    time (s) & $\tau_c$ & $\tau_t$ & $T_c$ & $T_t$\\\hline
    Model 1~\cite{Teusink2000} & $15.$ & $3.75$  & $339$ & $84.4$ \\    
    Model 2~\cite{Chassagnole2002} & $120$ &  $95.2$ & $2834$ & $2210$ \\
    Model 3~\cite{Mosca2012} & 4.94 &  0.16 & 107 & 3.53 \\
    Model 4~\cite{Curto1998} & $4.34~10^5$ &  $1.11~10^6$ & $9.35~10^6$ & $2.36~10^7$ \\    
  \end{tabular}
\caption{Value of the characteristic times $\tau_c$, $\tau_t$, $T_c$ and $T_t$ in
seconds for the four manually curated models~\cite{Curto1998, Teusink2000, Chassagnole2002, Mosca2012} we have studied and that
are available on the Biomodels repository~\cite{BioModels2006}.
}
\label{tab:biomodels_results}
\end{table}

\section{Discussion and conclusions}

The damping of a concentration fluctuation generally requires
the perturbation to spread out but in our reaction network
the drift plays a central role. The time scale for evacuating a perturbation
is what we call its lifetime $T$ 
(cf. Eqs.~\ref{eq:def_Tt} and ~\ref{eq:def_Tc}), though in other
contexts it can be referred to as the mean residence or transit time.
In the absence of a drift, corresponding to a pure diffusive regime,
the lifetime $T$ scales as the square of the diameter of the network,
a scaling which also arises for the standard measure of
return to equilibrium via the relaxation time $\tau$. That is the
situation one is most familiar with, and there $\tau$ provides
the longest characteristic time as it should.
However, for typical reaction networks, one has both diffusion and
drift. In particular, out of equilibrium systems will have fluxes,
and such fluxes may drive labeled atoms out of the network just
as drift does. In the presence of such drift, a perturbation's lifetime
$T$ can scale as the diameter of the network
divided by a characteristic drift velocity which is related to the
presence of flux. Interestingly, in this out of equilibrium situation,
the relaxation time $\tau$ is no longer informative about
the time scale of the (slow) process which evacuates perturbations. In particular,
in our toy model consisting of a homogeneous chain of reactions,
$\tau$ did not grow with the system size while $T$ grew linearly. We
showed analytically how that could be in that system, but the
phenomenon is general. Indeed, in the presence of drift, the linearized dynamics
can be decomposed into eigenvectors, but the leading eigenvector
determining $\tau$ tends to be concentrated on the metabolites
that can be excreted. As a result, $\tau$ is quite insensitive to the size
of the network while $T$ inevitably increases with network size since
the evacuation of a perturbation requires it to cross the diameter of the
network. These phenomena are most easily understood when the reactions
obey mass action, but they arise also for Michaelis-Menten-Henri reaction
laws. For this last case, the existence of a saturation of the flux
with concentration of metabolites exacerbates the difference between
$T$ and $\tau$. Interestingly, the dynamics of labeled atoms
that are often used to investigate kinetic properties of networks
are far less sensitive to these saturation effects. As a consequence,
the use of isotopic labelings can lead to severely underestimate
the longest characteristic time in these reaction networks.

\begin{acknowledgments}
We thank H. de Jong, D. de Vienne, C. Dillmann, J.-P. Mazat, and
D. Ropers for comments and references.
\end{acknowledgments}

\bibliographystyle{apsrev4-1}
\bibliography{aHenry2015.bib}


\end{document}